\documentclass[prl,twocolumn,superscriptaddress,amsmath,amssymb]{revtex4-1}
\usepackage{graphicx,bm}
\usepackage[colorlinks=true,citecolor=blue,urlcolor=blue]{hyperref}
\begin{document}
\title{Hybrid Quasi-Bound State in the Continuum at Topological Quantum Optics Interface}

\author{Yue-Zhi Zhang}
\affiliation{Center for Joint Quantum Studies and Department of Physics, School of Science, Tianjin University, Tianjin 300350, China}
\affiliation{Tianjin Key Laboratory of Low Dimensional Materials Physics and Preparing Technology, Tianjin University, Tianjin 300350, China}
\author{Leong-Chuan Kwek}
\affiliation{Centre for Quantum Technologies, National University of Singapore, 3 Science Drive 2, Singapore 117543}
\affiliation{MajuLab, CNRS-UNS-NUS-NTU International Joint Research Unit, Singapore UMI 3654, Singapore}
\affiliation{National Institute of Education, Nanyang Technological University, Singapore 637616}
\affiliation{Quantum Science and Engineering Centre, Nanyang Technological University, Singapore}
\author{Wei Nie}\email{weinie@tju.edu.cn}
\affiliation{Center for Joint Quantum Studies and Department of Physics, School of Science, Tianjin University, Tianjin 300350, China}
\affiliation{Tianjin Key Laboratory of Low Dimensional Materials Physics and Preparing Technology, Tianjin University, Tianjin 300350, China}

\begin{abstract}
Topological manipulation of light provides a versatile toolbox for photonic technologies. Here, we show that a topological atom array can induce photon localization in a waveguide via symmetry-protected light-matter interaction. Long-lived photon-atom entanglement reveals the existence of a novel topological quasi-bound state in the continuum (quasi-BIC). This hybrid light-matter quasi-BIC is formed at a critical coupling condition via collectively induced absorption, which is produced by quantum interference between edge and bulk states. We uncover the time-reversed relation between topological quasi-BIC and light amplification. Interestingly, one can realize a directional ultranarrow amplifier by means of critical coupling. Our work demonstrates an unconventional quasi-BIC at a topological quantum optics interface with potential applications in quantum devices.
\end{abstract}

\maketitle

\textit{Introduction}.---Bound states in the continuum (BICs) describe non-decaying discrete states whose energy levels lie in continuous spectrum of extended states~\cite{hsu2016,Kang2023,wang2024optical}. These exotic states have been realized in photonic systems via symmetry mismatch and destructive interference between decaying channels~\cite{PhysRevLett.100.183902,PhysRevLett.107.183901,hsu2013,PhysRevLett.113.037401,PhysRevLett.122.073601,PhysRevLett.130.093801,chen2023}. The ability to confine light makes BICs behave as optical cavities~\cite{PhysRevLett.119.243901}, and enables many applications, such as directional light emission~\cite{yin2020} and lasing~\cite{kodigala2017,huang2020ultrafast,yu2021,sang2022topological}. Photonic BICs usually have topological origins, e.g., vortex centers of polarization vectors in momentum space~\cite{PhysRevLett.113.257401,PhysRevLett.118.267401} and nondecaying edge states in real space~\cite{PhysRevLett.125.213901,hu2021nonlinear,wang2021quantum}, which inspire broad applications in photonic devices~\cite{jin2019topologically,wang2020generating,qin2025disorder}. In realistic systems, BICs suffer from environment-induced loss, producing quasi-BICs or high-Q supercavity modes with finite lifetime~\cite{sadrieva2017transition,PhysRevLett.121.193903,huang2023resonant}. Surprisingly, quasi-BICs are highly efficient in manipulating light, and yield a diverse range of photonic technologies~\cite{PhysRevLett.123.253901,PhysRevLett.126.073001,PhysRevApplied.16.064018,PhysRevLett.128.253901,zhang2022chiral,Schiattarella2024}.

Optical systems and gain media are prerequisites for light amplifiers, e.g., in cavity quantum electrodynamics (QED) with single atoms~\cite{McKeever2003,astafiev2007single}. Symmetry-protected topological photonic systems with gain can amplify optical fields via edge states or effective cavity modes~\cite{RevModPhys.91.015006,smirnova2020nonlinear}, and give rise to novel lasers~\cite{Jean2017lasing,bahari2017nonreciprocal,bandres2018topological,zhao2018topological,PhysRevLett.120.113901,zeng2020electrically,zhang2020low,yang2022}. Properties of topological amplifiers are determined by the interplay between gain and geometric structures of photonic lattices~\cite{PhysRevX.6.041026,PhysRevLett.122.143901}. In particular, one can realize
directional light amplification in topological photonic systems with tailored environments~\cite{PhysRevX.5.021025,wanjura2020,PhysRevA.103.033513,PhysRevB.103.L241408,PhysRevLett.127.213601,tian2023nonreciprocal}. Recently, optically driven topological matter exhibits strongly correlated quantum phenomena~\cite{bloch2022strongly,bao2022light}, including long-lived edge states~\cite{PhysRevLett.119.023603,PhysRevLett.124.083603}, topological phase transitions~\cite{PhysRevLett.115.045303,PhysRevLett.118.073602,PhysRevLett.127.250402,lin2023remote}, and vacuum-tuned quantum Hall effects~\cite{appugliese2022,PhysRevLett.131.196602,PhysRevX.15.021027,enkner2025tunable}. Quantum interfaces between light and topological matter might open a door to topology-enhanced quantum optical devices.

\begin{figure}[b]
\includegraphics[width=8.5cm]{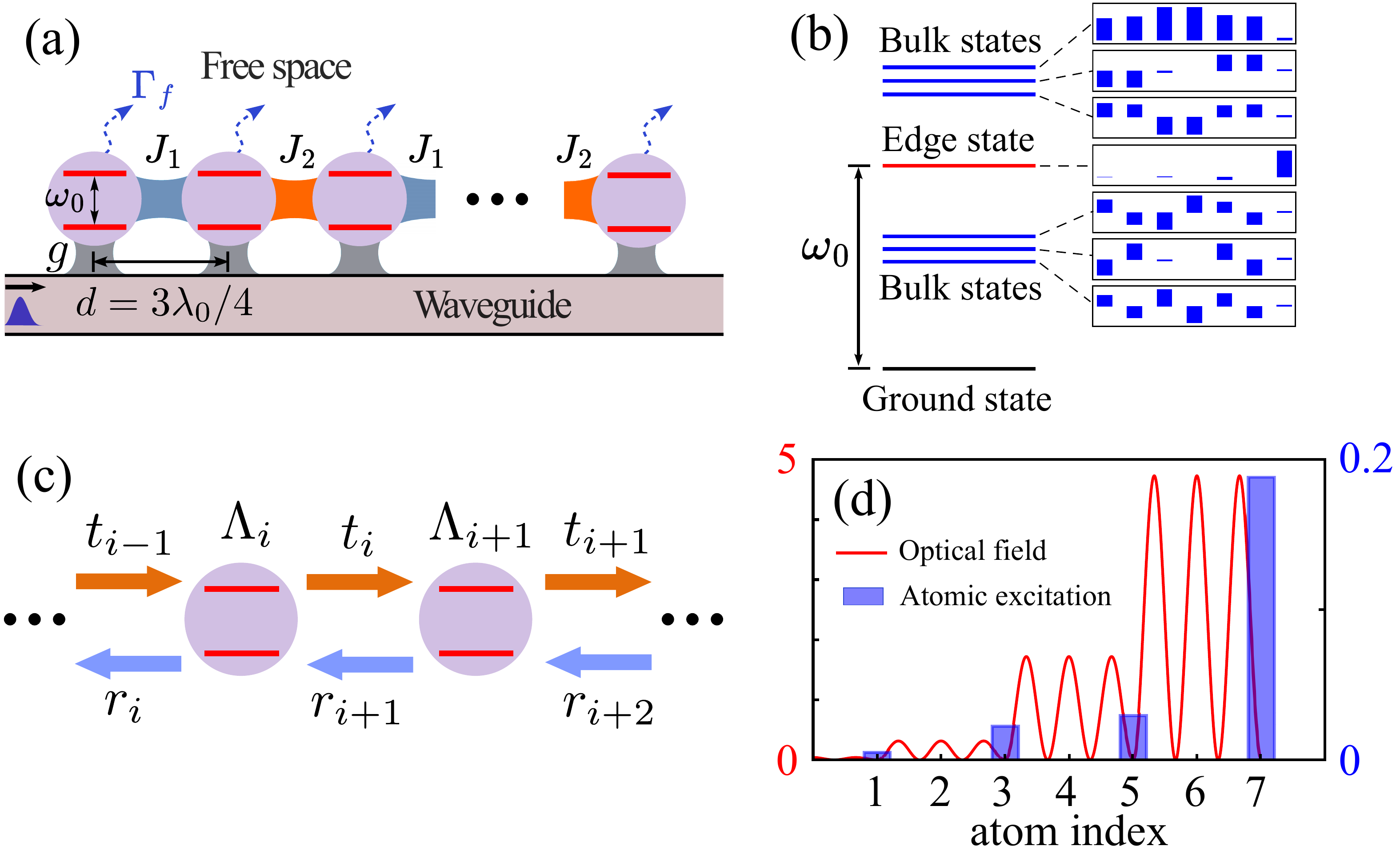}
\caption{(a) Schematics of a TAA-waveguide interface with photon-atom coupling $g$, atomic spacing $d$ and free-space dissipation $\Gamma_f$. (b) Energy spectrum of the TAA with a right edge state. (c) Atom-controlled photon transport with left-incident light. (d) Topology-induced photon localization. Both atomic excitation $|\lambda_i|^2$ and optical field intensity $|\phi(x)|^2$ are localized at the right boundary. We consider $J_0=2.2\Gamma$, $\varphi=0.2\pi$, $N=7$ in (b),(d); $\Gamma_f=0$ and $\epsilon=0.1$ in (d).}\label{fig1}
\end{figure}

In this work, we show that a quasi-BIC can be formed via \emph{symmetry-protected} light-matter interaction in waveguide QED with a topological atom array (TAA). It exhibits topologically protected photon-atom entanglement, different from photonic BICs or quasi-BICs. Optical properties of this hybrid light-matter quasi-BIC are nontrivially modified by loss and gain. Interestingly, time-reversed optical behaviors of the quasi-BIC give rise to directional amplification, without the requirement of tailored environments.

\textit{Topology-induced photon localization}.---A waveguide-TAA-coupled system is schematically shown in Fig.~\ref{fig1}(a). The Hamiltonian of the whole system is $H=H_\mathrm{TAA}+H_\mathrm{wg}+H_\mathrm{int}$. The TAA is described by the dissipative Su-Schrieffer-Heeger model~\cite{leykam2021probing} ($\hbar=1$) $H_\mathrm{TAA}=(\omega_0-i\Gamma_f)\sum_{i=1}^{N}\sigma_i^+\sigma_i^- + \sum_{i=1}^{N-1}J_i(\sigma_{i}^+ \sigma_{i+1}^- + \mathrm{H.c.})$, with atomic frequency $\omega_0$, weak free-space dissipation $\Gamma_f$, dimerized coherent couplings $J_i=J_0[1- (-1)^i\cos\varphi]$, and odd atom number $N$. Here, $\sigma_i^-=|g_i\rangle\langle e_i|$ and $\sigma_i^+=(\sigma_i^-)^\dagger$ are lowering and raising operators of two-level atoms. Energy spectrum and wave functions of the TAA are shown in Fig.~\ref{fig1}(b). A right-localized edge state is protected from bulk states. Hamiltonian of the waveguide is $H_{\mathrm{wg}}=-iv_{g}\sum_{\alpha=r,l}s_\alpha\int dx\hat{a}_\alpha^\dagger(x)\partial_x\hat{a}_\alpha(x)$, with photon group velocity $v_g$, $s_r=+1$ and $s_l=-1$. Annihilation (creation) operators for right- and left-propagating photons are $\hat{a}_{r,l}$ ($\hat{a}^{\dagger}_{r,l}$). Under Markovian and rotating-wave approximations, the light-matter interaction is $H_{\mathrm{int}}=g\int dx\sum_{i,\alpha} \delta(x-x_i)(\sigma_i^+\hat{a}_{\alpha}(x)+\text{H.c.})$, with photon-atom coupling $g$ and atomic positions $x_i$. Single-atom optical properties are determined by the decay rate $\Gamma=g^2/v_g$ ~\cite{PhysRevLett.95.213001,chang2007single}. Besides, we consider the atomic spacing $d=3\lambda_0/4$, with photonic wavelength $\lambda_0=2\pi v_g/\omega_0$.

Using the nonperturbative Bethe-ansatz approach~\cite{PhysRevA.76.062709}, one can exactly diagonalize the whole system within the single-excitation subspace. We consider a general state
\begin{equation}
    |\Psi(t)\rangle=\biggl[\sum_{\alpha=l,r}\int dx\phi_\alpha(x,t)\hat{a}_\alpha^\dagger(x)+\sum_{i=1}^{N}\lambda_i(t)\sigma_i^+\biggr]|0,g\rangle,
    \label{state}
\end{equation}
with the ground state $|0,g\rangle=|0\rangle\otimes_{i=1}^N|g_i\rangle$. Here, $|0\rangle$ denotes vacuum state in the waveguide, $\phi_{l,r}(x,t)$ are wave functions of left- and right-propagating photons, $\lambda_i(t)$ is excitation amplitude of the $i$th atom. The state in Eq.~\eqref{state} is complete because the TAA-waveguide Hamiltonian $H$ commutes with the total excitation operator $\mathcal{N}=\sum_i\sigma_i^+\sigma_i^-+\sum_\alpha\int dx\hat{a}_\alpha^\dagger(x)\hat{a}_\alpha(x)$. Considering a left-incident optical field with frequency $\omega$, the $i$th atom oscillates as $\lambda_i(t)=\Lambda_ie^{-i\omega t}$, and produces light transmission (reflection) amplitude $t_i$ ($r_i$), as shown in Fig.~\ref{fig1}(c). Substituting Eq.~\eqref{state} into Schr\"odinger equation $id_t|\Psi(t)\rangle=H|\Psi(t)\rangle$, we obtain equations of motion for photonic and atomic amplitudes
\begin{eqnarray}
i\partial_t\phi_\alpha &=&-iv_gs_\alpha\partial_x\phi_\alpha+g\sum_{i}\delta(x-x_i)\lambda_i,  \label{eom1}\\
id_t\lambda_i &=&(\omega_0-i\Gamma_f)\lambda_i +\delta_{i,\mathrm{odd}}(J_2\lambda_{i-1}+J_1\lambda_{i+1}) \notag\\
&&+\delta_{i,\mathrm{even}}(J_1\lambda_{i-1}+J_2\lambda_{i+1}) + g\sum_{\alpha}\phi_\alpha|_{x=x_i}. \label{eom2}
\end{eqnarray}
With ansatzes of propagating photons, one can obtain a set of linear algebraic equations
\begin{eqnarray}
\frac{1}{\sqrt{2\pi}}e^{ikx_i}(t_i-t_{i-1})&=&-i\frac{g}{v_g}\Lambda_i,     \label{algebraic1}\\
\frac{1}{\sqrt{2\pi}}e^{-ikx_i}(r_{i+1}-r_{i})&=&i\frac{g}{v_g}\Lambda_i.     \label{algebraic2}
\end{eqnarray}
Equations ~\eqref{eom1} and \eqref{eom2} describe photon-atom dynamics at the driven-dissipative TAA-waveguide interface. To pinpoint the role played by topology-protected light-matter interaction in manipulating optical fields, we firstly study steady state of the whole system with vanishing free-space dissipation ($\Gamma_f=0$). The driving field is a weak left-incident plane wave $\epsilon e^{i\omega x/v_g}/\sqrt{2\pi}$ with $\epsilon\ll 1$. In Fig.~\ref{fig1}(d), atoms at the right boundary are excited, and show symmetry-protected polarization, i.e., odd-site occupation~\cite{asboth2016short}. The optical field is weak in the waveguide away from the TAA; however, it is dramatically enhanced near the right boundary of TAA. Namely, the TAA-waveguide interface exhibits topology-enhanced photon localization. The localized optical field has discontinuous intensities in different sections of the waveguide separated by polarized atoms. The reason is that atoms serve as potential barriers and change optical field distribution in the array, as shown by Eqs.~\eqref{algebraic1} and \eqref{algebraic2}. The pattern of photon localization can be tuned by TAA.

\begin{figure}[b]
\includegraphics[width=8.5cm]{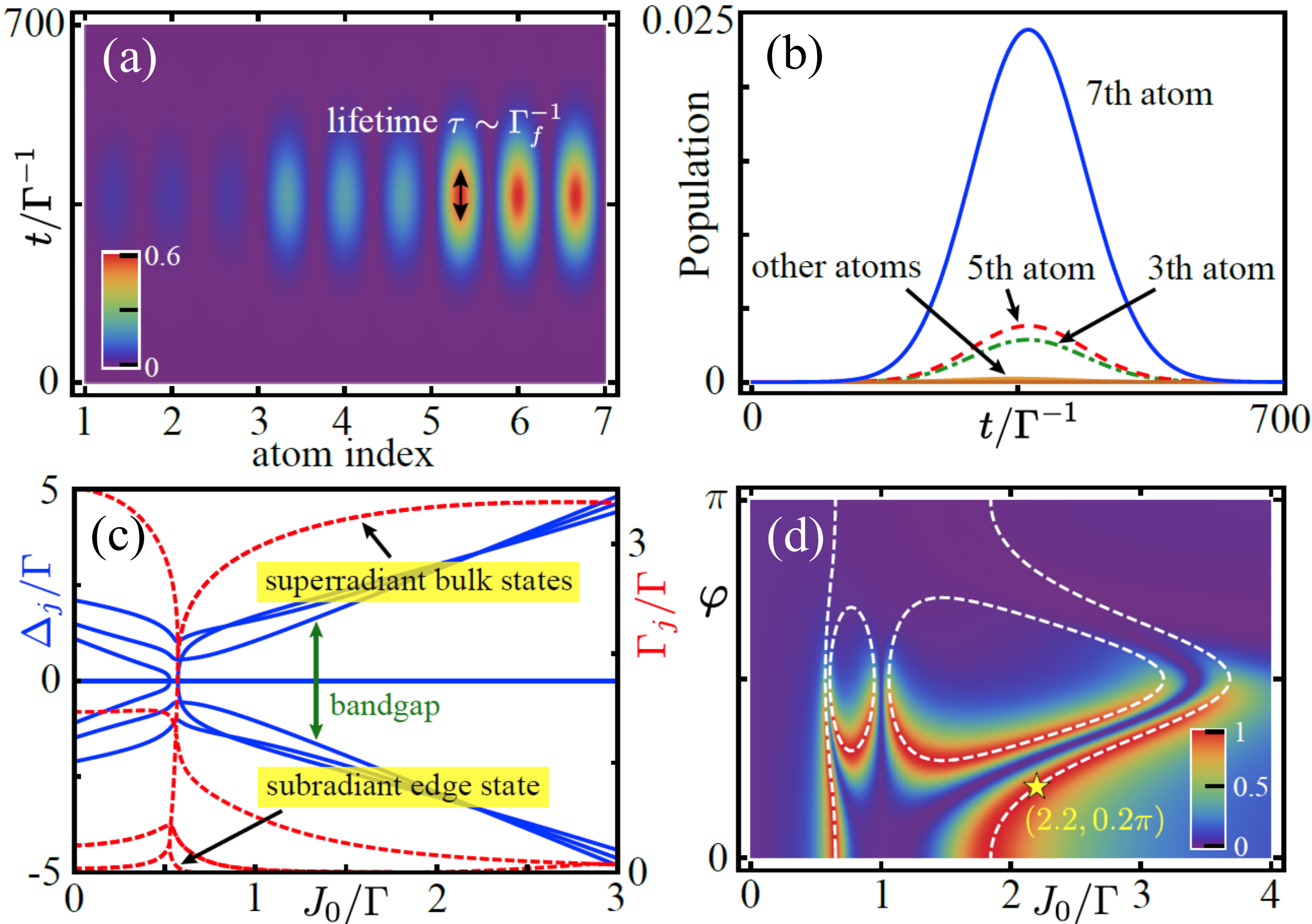}
\caption{Dynamics of the hybrid quasi-BIC with (a) photon intensity $|\phi(x,t)|^2$ and (b) atomic excitation $|\lambda_i(t)|^2$. (c) Relative frequencies (blue-solid) and decay rates (red-dashed) of collective eigenmodes. At $(J_0,\varphi)=(2.2\Gamma,0.2\pi)$, decay rate of the edge state is $\Gamma_{\mathrm{edge}}\approx 0.013\Gamma$. (d) Photon absorption at $\omega=\omega_0$ changed by coupling parameters. Absorption reaches the maximum at $\Gamma_{\mathrm{edge}}=\Gamma_f$ (white-dashed curves) for $\varphi<\pi/2$. We consider $J_0=2.2\Gamma$ in (a),(b); $\varphi=0.2\pi$ in (a)-(c); $\Gamma_f=0.013\Gamma$ in (a),(b),(d); $N=7$ in (a)-(d).}
\label{fig2}
\end{figure}

\textit{Loss-induced quasi-BIC with topology-protected light-matter interaction}.---Free-space dissipation $\Gamma_f$ plays an important role in controlling photon-atom dynamics of the TAA-waveguide-coupled system. Figure~\ref{fig2}(a) shows photon dynamics induced by a TAA with weak free-space dissipation. We consider the driving of a single-photon Gaussian wave packet. Optical field localized in the TAA gradually strengthens, and decreases after reaching the maximum. Crucially, the time of photon dynamics depends on the free-space dissipation $\tau \sim \Gamma_f^{-1}$. Edge atoms display excitation dynamics synchronized with localized photon, as shown in Fig.~\ref{fig2}(b). Therefore, the single-photon pulse excites a photon-atom entangled state with long lifetime, i.e., a hybrid light-matter quasi-BIC. Waveguide QED with small $\Gamma_f$ ($\sim 0.01\Gamma$) has been experimentally realized in superconducting quantum circuits~\cite{mirhosseini2019cavity}. This loss-induced optically drivable quasi-BIC is distinct from dark states that are decoupled from waveguide~\cite{RevModPhys.95.015002}.

To understand the generation of hybrid quasi-BIC, we trace out photonic degrees of freedom for weak atom-waveguide coupling, and obtain a master equation~\cite{caneva2015quantum}
\begin{equation}
    d_t\rho=-i[(H_\mathrm{eff}+H_{\mathrm{d}})\rho-\rho (H_\mathrm{eff}+H_{\mathrm{d}})^\dagger]+2\Gamma\sum_{i,j}\sigma_{i}^-\rho\sigma_{j}^+,
\end{equation}
with an effective non-Hermitian Hamiltonian $H_\mathrm{eff}=H_{\mathrm{TAA}} -i\Gamma\sum_{i,j}e^{ik_0|x_i-x_j|}\sigma_i^+\sigma_j^-$. The second term in $H_\mathrm{eff}$ denotes waveguide-induced atom-atom couplings~\cite{gonzalez2024light}. The driving Hamiltonian is $H_\mathrm{d}=\epsilon g\sum_i (\sigma_i^+ e^{i k x_i -i\omega t} +\mathrm{H.c.})$. We diagonalize the non-Hermitian Hamiltonian as $H_\mathrm{eff}=\sum_j (E_j-i\Gamma_f)|\psi_j^R\rangle\langle\psi_j^L|$. These eigenmodes are collective quantum states, i.e., subradiant (superradiant) states with inhibited (enhanced) dissipation~\cite{RevModPhys.95.015002}, compared to the single-atom decay rate $\Gamma$. The topological phase of TAA survives for the atomic spacing $d=3\lambda_0/4$, because of preserved chiral symmetry~\cite{PhysRevLett.127.250402}. In Fig.~\ref{fig2}(c), we show relative frequencies $\Delta_j=\mathrm{Re}(E_j)-\omega_0$ and decay rates $\Gamma_j=-\mathrm{Im}(E_j)$ of collective eigenmodes. Zero frequency shift of the waveguide-dressed edge state implies symmetry-protected photon-atom coupling in $H_{\mathrm{int}}$. Particularly, in the topological phase with large bandgap, there exist a subradiant edge state with small decay rate $\Gamma_{\mathrm{edge}}$ and two superradiant bulk states. Quantum interference in optical responses of collective edge and bulk states is controlled by the bandgap.

With the input-output method~\cite{caneva2015quantum}, transmitted and reflected fields through the TAA are
\begin{eqnarray}
\hat{a}_t(x) &=& \epsilon e^{i k x}-i\sum_i \sqrt{\frac{\Gamma}{v_g}}\sigma_i^- e^{i k_0(x-x_i)}, \\
\hat{a}_r(x) &=& -i\sum_i \sqrt{\frac{\Gamma}{v_g}}\sigma_i^- e^{i k_0(x+x_i)},
\end{eqnarray}
respectively. The transmission (reflection) amplitude is $t(r)=\langle \hat{a}_{t(r)}(x) \rangle/\epsilon e^{i k x}$. Photon absorption is $\eta=1-T-R$, with $T=|t|^2$ and $R=|r|^2$. The free-space dissipation is responsible for photon absorption. A single atom has maximal photon absorption $\eta=0.5$ at $\Gamma_f=\Gamma$~\cite{PhysRevLett.102.173602}. This absorption can be achieved in atom arrays with small $\Gamma_f$ ($\Gamma_f\ll \Gamma$) via collectively induced absorption (CIA)~\cite{cheng2024collectively}, because of destructive quantum interference between subradiant and superradiant states. Interestingly, topology-protected light-matter interaction can further boost photon absorption with small $\Gamma_f$. In Fig.~\ref{fig2}(d), we show photon absorption at resonant driving. Topology-enhanced CIA with $\eta\approx 1$ corresponds to $\Gamma_{\mathrm{edge}}=\Gamma_f$. This explains the lifetime $\tau \sim \Gamma_f^{-1}$ of the hybrid quasi-BIC. Moreover, absorption of left-incident photon is enhanced for the TAA with a right-localized edge state, i.e., $\varphi<\pi/2$, due to nonreciprocal optical response of the edge state in reflection process~\cite{PhysRevApplied.15.044041}. The condition $\Gamma_f=\Gamma_{\mathrm{edge}}$ of topology-enhanced CIA for generating the hybrid quasi-BIC is \emph{critical coupling}, analogous to whispering-gallery-mode microcavities~\cite{yariv2000,PhysRevLett.85.74,Xiao2020Microcavities} and topological cavities~\cite{yu2021critical,kumar2022active}.

\begin{figure}[t]
\includegraphics[width=8.5cm]{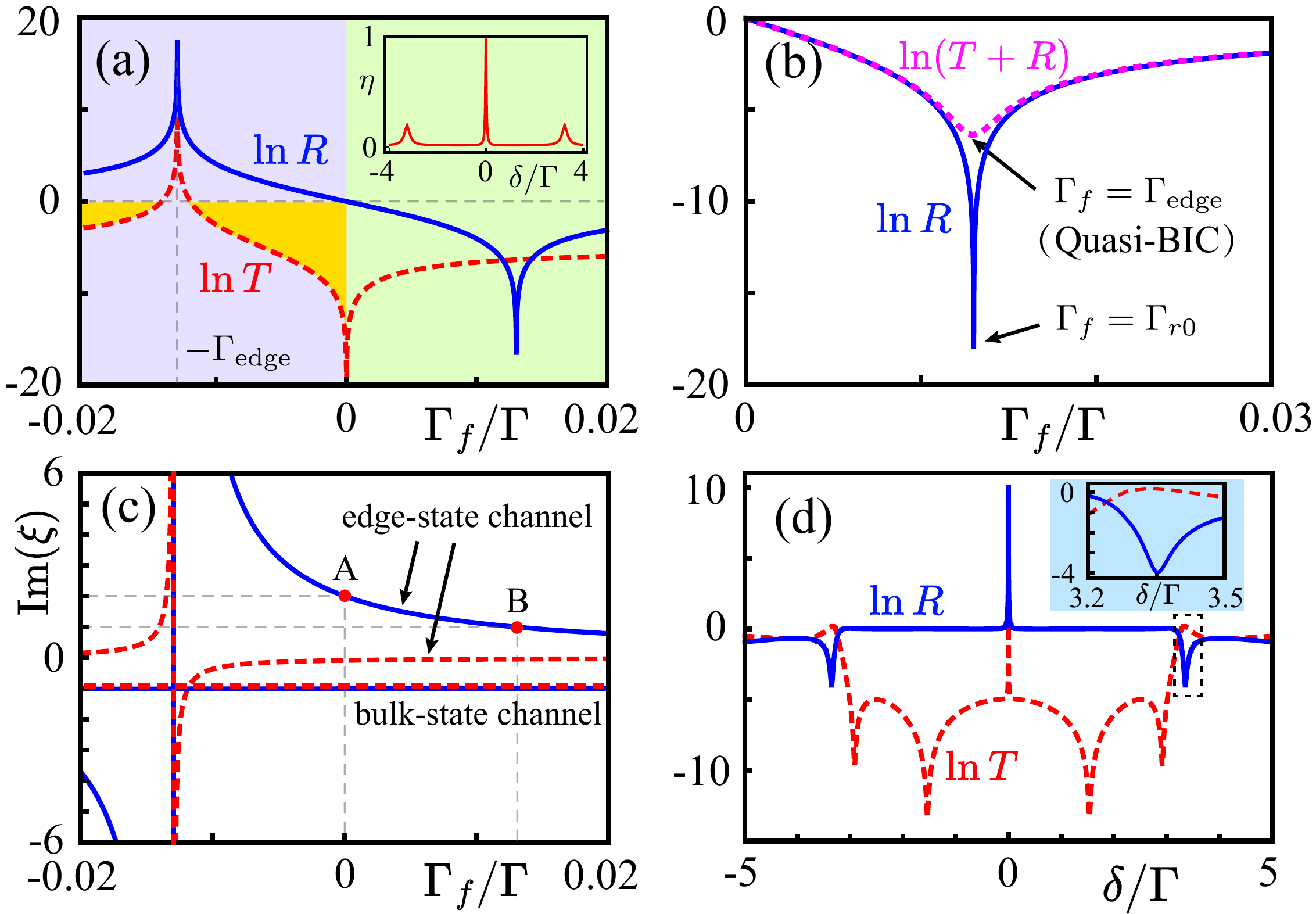}
\caption{(a) Logarithms of transmission (red-dashed) and reflection (blue-solid) in loss ($\Gamma_f>0$) and gain ($\Gamma_f<0$) regimes. Light transmission through the TAA with gain can be damped (yellow regime). The inset shows CIA versus light-atom detuning $\delta$ at critical coupling. (b) Logarithms of reflection (blue-solid) and the sum of transmission and reflection (purple-dotted) are minimized at $\Gamma_f=\Gamma_{r0}$ and $\Gamma_f=\Gamma_{\mathrm{edge}}$, respectively. (c) Quantum interference between edge- and bulk-state channels tuned by $\Gamma_f$. $\mathrm{Im}(\xi)$ for transmission (reflection) process is denoted by red-dashed (blue-solid) curve. At point A (B), $\mathrm{Im}(\xi_{\mathrm{edge}}^r)$ is $2$ ($1$). (d) Ultranarrow light amplification at $\Gamma_f=-\Gamma_{\mathrm{edge}}$. The inset shows weak transmissive amplification induced by subradiant bulk states. Other parameters are the same as Fig.~\ref{fig1}.}\label{fig3}
\end{figure}

\textit{Time-reversed topological quasi-BIC}.---Photon absorption and amplification are time reversal counterparts~\cite{PhysRevLett.105.053901,PhysRevA.82.031801}. For example, time-reversed lasing enables coherent perfect absorption (CPA)~\cite{wan2011time,wong2016lasing,baranov2017}. Here, we study light amplification via time reversal of topology-enhanced CIA. The time reversal can be implemented by changing the free-space dissipation to gain ($\Gamma_f<0$). In Fig.~\ref{fig3}(a), we show photon transmission and reflection for a left-incident optical field with frequency $\omega=\omega_0$. At $\Gamma_f=0$, light is completely reflected, with vanishing transmission. In the loss regime ($\Gamma_f>0$), photon transmission and reflection are inhibited. Particularly, $\ln (T+R)$ and $\ln R$ are minimized at $\Gamma_f=\Gamma_{\mathrm{edge}}$ and $\Gamma_f=\Gamma_{r0}$, respectively, as shown in Fig.~\ref{fig3}(b). Because of $\Gamma_f$-dependent transmission, $\Gamma_{r0}$ is slightly different from $\Gamma_{\mathrm{edge}}$. The inset in Fig.~\ref{fig3}(a) shows CIA at critical coupling. Topology-enhanced ultranarrow CIA of the hybrid quasi-BIC reveals the characteristic of high-quality cavity~\cite{armani2003ultra}. Two off-resonant absorption peaks are induced by bulk states. In the gain regime ($\Gamma_f<0$), reflected light shows amplification even for weak gain. However, transmitted light is not magnified until near $\Gamma_f=-\Gamma_{\mathrm{edge}}$. At the time-reversed critical coupling, ultimate amplification can be realized in both reflection and transmission processes. Importantly, the quasi-BIC releases the requirement of large gain for light amplification in waveguide QED systems~\cite{PhysRevLett.104.183603,PhysRevLett.110.263601,PhysRevLett.120.063603}.

To further clarify the mechanism of light amplification, we use the multichannel scattering method~\cite{PhysRevApplied.15.044041}. Reflection and transmission amplitudes are $r=-i\sum_j \Xi_j/\varepsilon_j$, $t=1-i\sum_j \tilde{\Xi}_j/\varepsilon_j$, respectively. Here, we have $\varepsilon_j=\delta-\Delta_j+i(\Gamma_j+\Gamma_f)$, with $\delta=\omega-\omega_0$. Light-matter interactions in photon scattering processes are characterized by interaction spectra $\Xi_j=\mathbf{V}^{\mathrm{T}}|\psi_j^R\rangle\langle\psi_j^L|\mathbf{V}$ and $\tilde{\Xi}_j=\mathbf{V}^{\dagger}|\psi_j^R\rangle\langle\psi_j^L|\mathbf{V}$, with $\mathbf{V}=(e^{s_{\alpha} ik_0x_1},e^{s_{\alpha} ik_0x_2},\ldots,e^{s_{\alpha} ik_0x_N})^{\mathrm{T}}$. At resonant driving ($\delta=0$), interaction spectra for the edge state are determined by its dissipation. For simplicity, we consider photon reflection amplitudes $\xi^r_{\mathrm{edge}}=\Xi_{\mathrm{edge}}/\varepsilon_{\mathrm{edge}}$, $\xi^r_{\mathrm{bulk}}=\sum_{j\in \mathrm{bulk}} \Xi_j/\varepsilon_j$ by the edge- and bulk-state channels, respectively. Similarly, we can define $\xi^t_{\mathrm{edge}}$ and $\xi^t_{\mathrm{bulk}}$ for transmission process.

The free-space dissipation nontrivially tunes quantum interference between different scattering channels. In Fig.~\ref{fig3}(c), we show $\Gamma_f$-tuned $\mathrm{Im}(\xi)$ for edge- and bulk-state channels in transmission and reflection processes. For resonant driving, bulk-state channel is insensitive to $\Gamma_f$ due to the bandgap. However, edge-state channel is significantly changed by $\Gamma_f$ due to its subradiant nature, modifying edge-bulk quantum interference. At $\Gamma_f=0$ (point A) and $\Gamma_f=\Gamma_{r0}$ (point B), destructive quantum interferences between edge- and bulk-state channels produce vanishing transmission and reflection, respectively. The edge-bulk quantum interference enhances photon absorption at $\Gamma_f=\Gamma_{\mathrm{edge}}$. At the time-reversed critical coupling $\Gamma_f=-\Gamma_{\mathrm{edge}}$, light is amplified in transmission and reflection processes because of divergent optical responses at $\varepsilon_{\mathrm{edge}}=0$. As shown in Fig.~\ref{fig3}(d), transmitted and reflected photons are magnified at resonance. The inset shows reduced reflection and weakly amplified transmission induced by bulk states.

\begin{figure}[t]
    \includegraphics[width=8.5cm]{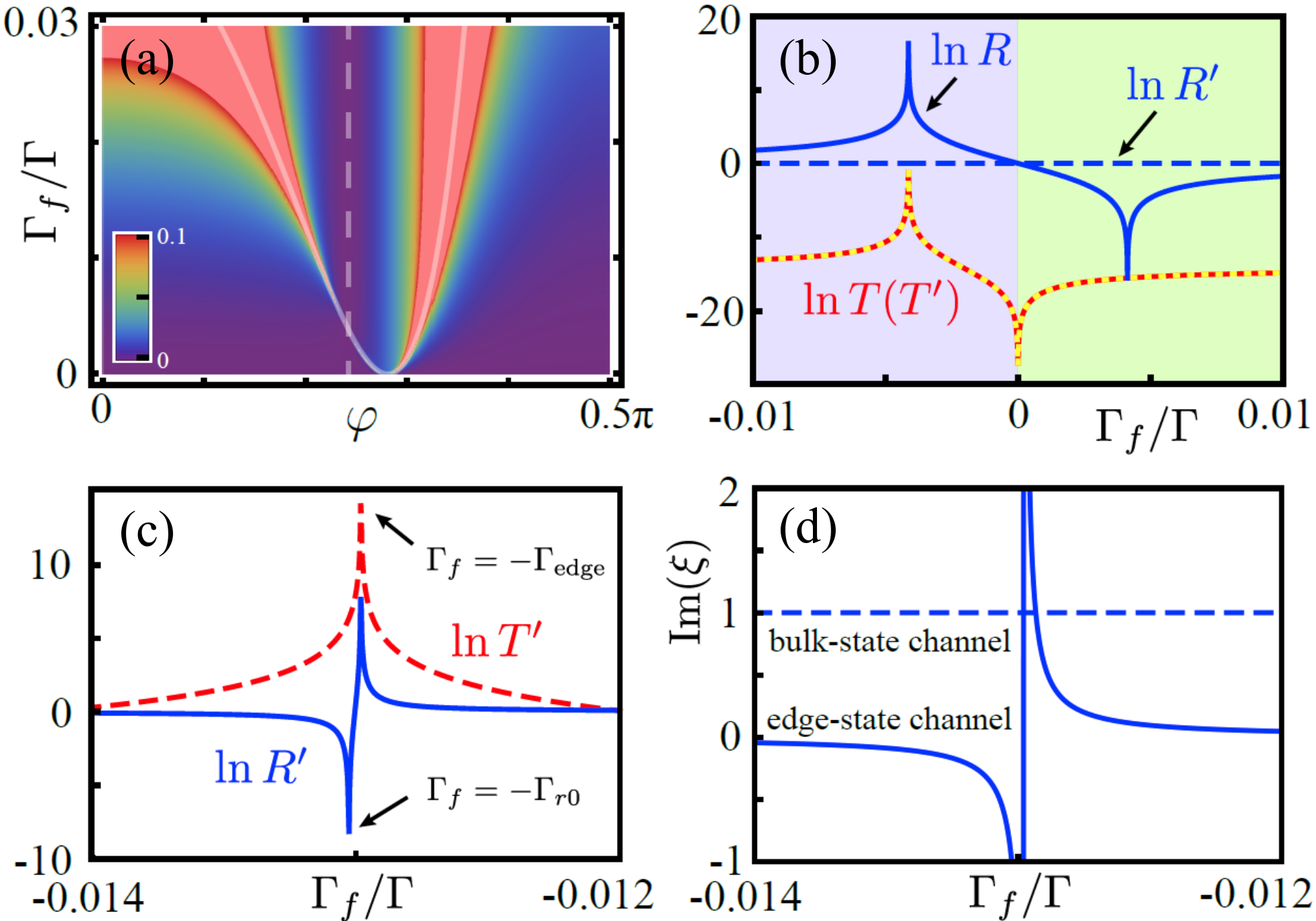}
    \caption{(a) Time-reversal property $\delta\chi$ of reflection for left-incident light. The white-dashed line indicates $\Gamma_{\mathrm{edge}}=\Gamma_{r0}$ around $\varphi= 0.241\pi$ with time-reversed reflection ($\delta\chi=0$). The white-solid curve shows $\Gamma_f=\Gamma_{\mathrm{edge}}$ as a function of $\varphi$. Pink regime denotes $\delta\chi>0.1$. (b) Photon transport at $\varphi=0.241\pi$. $T$ ($R$) and $T'$ ($R'$) denote transmission (reflection) of left- and right-incident light, respectively. (c) Transmissive amplification at $\varphi=0.2\pi$. (d) Reflectional amplitudes for edge and bulk channels. Other parameters are the same as Fig.~\ref{fig1}.}
    \label{fig4}
\end{figure}

\textit{Critical coupling vs directional amplification}.---In the TAA-waveguide-coupled system, reflection process has nontrivial time-reversed properties near critical coupling. We define light reflection at resonance as $R_0=|r(\delta=0)|^2=e^{\chi(\Gamma_f)}$, where $\chi(\Gamma_f)$ describes optical amplification or damping in reflection process. Thus, the parameter $\delta\chi=|\chi(\Gamma_f)+\chi(-\Gamma_f)|$ characterizes time-reversal property of optical response. Time-reversed reflection, i.e., $\chi(-\Gamma_f)=-\chi(\Gamma_f)$, can be realized at $\Gamma_{\mathrm{edge}}=\Gamma_{r0}$. This means that $\partial T/\partial \Gamma_f \approx 0$, such that $T+R$ and $R$ simultaneously reach minimum at the same $\Gamma_f$. In Fig.~\ref{fig4}(a), we demonstrate $\delta\chi$ for the left-incident light. The reflection is not time-reversed for $\Gamma_{\mathrm{edge}}\neq\Gamma_{r0}$ near the critical coupling (white-solid curve), with an exception at $(J_0,\varphi)=(2.2\Gamma,0.241\pi)$ where time-reversed reflection is achieved in a broad range of $\Gamma_f$. In Fig.~\ref{fig4}(b), we show transmission and reflection for left- and right-incident light at these two coupling parameters. Reflection process is time-reversed ($\delta \chi=0$) for both left- and right-incident light. In particular, photon reflection is nonreciprocal ($R\neq R'$), and the left-incident light has ultimate amplification at $\Gamma_f=-\Gamma_{\mathrm{edge}}$. Transmission process is reciprocal ($T=T'$) and is inhibited in the gain regime. Therefore, reflectional amplification of left-incident light can be realized. Broken time-reversal reflection ($\delta\chi>0$) in Fig.~\ref{fig4}(a) also indicates interesting optical response. In Fig.~\ref{fig4}(c), we show photon transmission and reflection for the right-incident light in gain regime. Ultimate amplification is achieved for both transmission and reflection processes at the time-reversed critical coupling. Surprisingly, by slightly increasing the gain, the amplified reflection becomes damped with the minimum at $\Gamma_f=-\Gamma_{r0}$.

The gain-induced amplification-damping transition in reflection is attributed to topology-protected edge state. For $\Gamma_f +\Gamma_{\mathrm{edge}}>0$, bulk- and edge-state channels have constructive quantum interference in reflection, as shown in Fig.~\ref{fig4}(d). $\mathrm{Im}(\xi)$ of bulk-state channel is positive for the right-incident light, different from left-incident light (see Fig.~\ref{fig3}(c)). For $\Gamma_f +\Gamma_{\mathrm{edge}}<0$, the edge-state channel produces a $\pi$ phase shift of reflected light. Moreover, the interaction spectra $\tilde{\Xi}_{\mathrm{edge}}$ is small, due to subradiance of the edge state. Accordingly, a slight change of optical gain leads to vanishing reflection via destructive edge-bulk quantum interference. Time-reversal breaking of reflection near critical coupling enables transmissive amplification. Therefore, the hybrid quasi-BIC produces a tunable, directional and ultranarrow amplifier, which would be useful for waveguide-integrated quantum devices.

\textit{Conclusions}.---In this work, we uncover a novel quasi-BIC via interaction between light and dissipative topological matter. Symmetry-protected light-matter interaction in the TAA-waveguide-coupled system gives rise to the critical coupling, in which a quasi-BIC can be generated with weak free-space dissipation. Different from conventional photonic BICs or quasi-BICs, this hybrid light-matter quasi-BIC shows spatiotemporal photon-atom entanglement. The time-reversed relation between topology-enhanced CIA and photon amplification makes the quasi-BIC behave as an artificial cavity, in which optical fields can be efficiently amplified with weak gain. Topology-controlled collective quantum properties of edge and bulk states near time-reversed critical coupling enable directional photon amplification. Our work presents a theory of hybrid topological quasi-BIC, and may shed new light on quantum optics of topological matter.

\begin{acknowledgments}
The authors thank Daniel Leykam for a critical reading and insightful comments. W.N. is supported by the National Natural Science Foundation of China (Grants No. 92476115 and No. 12105025). L.-C.K. acknowledges support from the Ministry of Education, Singapore, and the National Research Foundation, Singapore.
\end{acknowledgments}

\end{document}